\documentclass[final, 11pt, twocolumn]{article}
\usepackage[table,xcdraw]{xcolor}
\usepackage{neurips_2024}

\usepackage[utf8]{inputenc} 
\usepackage[T1]{fontenc}    
\usepackage{hyperref}       
\usepackage{url}            
\usepackage{booktabs}       
\usepackage{amsfonts}       
\usepackage{nicefrac}       
\usepackage{microtype}      
\usepackage{orcidlink}
\usepackage{cuted}
\usepackage{bbm}
\usepackage{multirow}
\usepackage{float}
\usepackage{amssymb}
\usepackage{amsmath}
\usepackage{bm}
\usepackage[square,numbers]{natbib}
\usepackage[toc,page]{appendix}

\title{Equivariant Machine Learning Decoder for 3D Toric Codes}

\author{%
  Oliver Weissl \orcidlink{0009-0008-7575-0187}\\
  \texttt{weissl@fortiss.org} \\
  \And
  Evgenii Egorov \orcidlink{0000-0003-1434-7817}\\
  AMLab - University of Amsterdam \\
  \texttt{egorov.evgenyy@ya.ru} \\
}

\begin{document}
\maketitle

\begin{abstract}
\begin{quotation}
      Mitigating errors in computing and communication systems has seen a great deal of research since the beginning of the widespread use of these technologies. However, as we develop new methods to do computation or communication, we also need to reiterate the method used to deal with errors. Within the field of quantum computing, error correction is getting a lot of attention since errors can propagate fast and invalidate results, which makes the theoretical exponential speed increase in computation time, compared to traditional systems, obsolete. To correct errors in quantum systems, error-correcting codes are used. A subgroup of codes, topological codes, is currently the focus of many research papers. Topological codes represent parity check matrices corresponding to graphs embedded on a $d$-dimensional surface. For our research, the focus lies on the toric code with a 3D square lattice. The goal of any decoder is robustness to noise, which can increase with code size. However, a reasonable decoder performance scales polynomially with lattice size. As error correction is a time-sensitive operation, we propose a neural network using an inductive bias: equivariance. This allows the network to learn from a rather small subset of the exponentially growing training space of possible inputs. In addition, we investigate how transformer networks can help in correction. These methods will be compared with various configurations and previously published methods of decoding errors in the 3D toric code.
\end{quotation}
\end{abstract}


\section{Introduction}
\label{sec:introduction}
Errors are a part of most communication and computation systems imaginable; however, these errors are destructive and unwanted, therefore methods to mitigate them are required. This necessity led to the field of error correction, which essentially encodes a message or data with redundancy to make accurate reconstruction possible, even with an error present. Challenges in this field form around the size of redundancy needed and the threshold of how many errors can be corrected without the system falling into a "chaotic" uncorrectable state. In traditional computational and communication systems, the error would be expressed as a bit switch, which can be identified with enough certainty, using classical error correction strategies. There are many ways to achieve this in bit-based systems, such as the Hamming code \cite{hamming1950error}, or parity check codes.
Parity check codes encode information (logicals) in redundant physical bits, facilitating parity checks.

Decoders take violations of parity checks as input; however, they can have either a low- or high-level approach to dealing with these errors.
In a low-level decoder, the aim is to reconstruct the physical noise acting on the physical qubits, to correct the error. The high-level decoder, on the other hand, aims to predict the logical information from the input. An approach to the high-level decoder is a maximum likelihood decoder, which finds the most likely logical information, based on the input received.

Although this problem of correcting errors has received a lot of attention in classical communication and computation systems, with the emergence of more powerful quantum computing approaches in recent years, the domain of error correction in quantum computing and communication has grown significantly \cite{GYONGYOSI201951}. In addition to traditional systems, quantum systems suffer from more noisiness, due to the fragile nature of quantum information. This amplifies the need for error correction strategies, as only then can the potential superiority to traditional computational systems be built upon \cite{gambetta2017building}.

Contrary to the classical computing approach, we have to deal with not only bit-flip errors but also two different kinds of errors.

Errors in the error correction code are the result of an operator string acting on the physical qubits. Concretely, we consider X (bit-flip) and Z (phase-flip) errors as operators, which can be represented using Pauli matrices (\ref{eq:pauli}).
\begin{equation}\label{eq:pauli}
\begin{aligned}
    X = 
\begin{bmatrix}
0 & 1 \\
1 & 0
\end{bmatrix}\quad, \quad
    Z = \begin{bmatrix}
1 & 0 \\
0 & -1
\end{bmatrix}.\\
\end{aligned}
\end{equation}
These errors show different manifestations when applied to a qubit $|\psi\rangle$ with superposition $\alpha|0\rangle + \beta|1\rangle$ (\ref{eq1}) \cite{lidar2013quantum}. 
\begin{equation} \label{eq1}
\begin{split}
X(|\psi\rangle) & = \beta|0\rangle + \alpha|1\rangle,\\
Z(|\psi\rangle) & = \alpha|0\rangle - \beta|1\rangle.
\end{split}
\end{equation}
To detect those errors, parity checks are employed; similarly to the classical computing approaches, these quantum parity checks can be constructed by encoding logical qubits into multiple physical qubits. With this encoding, information can be extracted, allowing the detection of physical errors. However, due to the additional commutativity restrictions for parity checks in the quantum case, the degeneracy of code space and hence decoding complexity increases exponentially with the number of physical qubits. 

A particular class of these quantum codes is that of topological codes. Topological codes essentially encode logical qubits with multiple physical qubits on a discrete lattice that tessellates an n-dimensional manifold.
One such topology construction introduced by Kitaev \cite{kitaev2003fault} goes as follows: physical qubits are organized on the edges of Tanner graphs, which are embedded in a $d$-dimensional torus manifold. This construction results in the increasingly prominent toric code \cite{kitaev2003fault}. The faces and dual faces of the graph are used to organize parity checks, with the fundamental group of the torus encompassing all logical operators \cite{kitaev2003fault}. In the case of the square lattice, faces are tiling the manifold, with dual faces expressing themselves as the co-boundaries of edges on the primal.

Following from the previous description of the logic behind quantum codes, we can see that the difficulty of optimally reconstructing the errors grows exponentially with the size of the code; in fact, it is determined to be complete $\#$ P \cite{7097029}. This extremely large search space for tracing back errors forces decoders to be based on rule-based algorithms (non-trainable) or some machine learning architecture (trainable), which can generalize well in the case of topological codes. The drawback of heuristic-based approaches is the inflexibility concerning the topology and size of the quantum codes. For solutions to be more generalizable, neural decoders are potentially powerful options \cite{egorov2023end}. 

Previous research on the 2D toric code has shown multiple possible decoder approaches, such as the minimum weight perfect matching decoder (MWPM)\cite{dennis2002topological} or the union find decoder \cite{Delfosse2021almostlineartime}.
Although these approaches are fast and perform decently, they have several major flaws; they treat X and Z errors independently, not considering their interactions. Additionally, they do not consider the degeneracy of the toric code, which forms due to cycle-invariant parity checks. In other words, this means that parity checks can produce the same syndrome for different error paths. Reconstructing this path or predicting the affected logicals is the task of the decoder, which is less likely to be accurate if it does not consider the inherent symmetries and biases originating from the previously elaborated characteristics.

This is where neural decoders can excel. Although the possible error space is large, we can leverage symmetries within the lattice to simplify and accelerate decoding. Egorov et al. established the efficiency of machine learning decoders with equivariance properties in the task of decoding two-dimensional toric code to depolarize noise \cite{egorov2023end}.
In this project, we want to extend those equivariance properties of neural decoders into 3D toric code. Our research questions are as follows.

\begin{itemize}
    \item How can the equivariance properties of the maximum likelihood decoder for graphs embedded in 3-dimensional manifolds be described?
    \item How can those equivariance constraints be embedded in a neural decoder architecture?
    \item How effective is the proposed neural decoder with respect to other non-trainable and trainable decoders? (Specifically, the logical accuracy of the decoder)?
\end{itemize}


\section{Related Work}
\label{sec:related_work}
In current research, most error correction strategies focus on 2D code, while extending to 3 dimensions would allow systems to leverage more transverse logical gates and have a theoretically higher fault tolerance overall \cite{PhysRevA.100.012312}. To achieve universal quantum computation, a certain set of gates must be available, which can be expressed as any set of substitutes that includes gates of Clifford hierarchy 3 and higher \cite{gottesman1999demonstrating, anderson2022groups}. As proved by Bravyi et al., every surface code of n dimensions includes gates of the n-th Clifford hierarchy, making 3D surface codes capable of facilitating universal quantum computation \cite{Bravyi_2013}.

Non-trainable decoders for the 3D Toric code, such as SweepMatch \cite{Vasmer_2021} or BP-OSD \cite{panteleev2021degenerate}, showcase promising thresholds; however, only benchmark Phase-Flip errors. These approaches exploit the fact that decoding Phase-Flip errors can be easily adapted to decode bit-flips as well; however, they require multiple shots for error correction to work. This dependency on running multiple error correction cycles is a limiting factor, as neural decoders such as TheEND \cite{egorov2023end}, have shown good performance in one-cycle decoding.

Current approaches to neural decoders show great potential to exploit symmetry properties in topological codes. Egorov et al. argues in their work that pure neural decoder approaches suffer from scalability, while combining them with heuristics limits their potential accuracy. Therefore, they proposed the usage of equivariant neural decoders. As shown in their work, these geometric models perform well in generalizing errors due to exploiting symmetries observed in the 2D toric surface code \cite{egorov2023end}. Although this approach works well on 2D surface codes, the extension of such principles into 3D is more challenging, since X and Z errors form in additional dimensions on 3D codes. 

This mismatch in dimensionality also prohibits us from using classical decoders such as MWPM on both X- and Z-syndromes. Although these decoders work well on the 2D implementation of topological codes, they can only be used to reconstruct Z errors in 3D topological codes, as argued by Quintavalle et al. \cite{Quintavalle_2021}. In their work, they investigated two methods of error correction for 3D toric codes, using two-stage error correction. The first step entails correcting the syndrome noise, with the following step correcting the respective errors on the qubits. Both solutions use belief propagation + ordered statistics decoding (BP-OSD \cite{fossorier2001iterative, panteleev2021degenerate}) as their second stage, while for the first stage, they have one implementation using MWPM and another implementation using BP-OSD as well \cite{Quintavalle_2021}. Although in their research they achieve a promising error threshold of around 2.9\%, the question of whether exploiting potential symmetries in the 3D code geometry is possible remains unanswered. Additionally, it should be noted that BP-OSD can hardly be parallelized due to the serial nature of OSD\cite{fossorier1995soft}, therefore the computational complexity is high \cite{fossorier2001iterative}. 

One way to capture the information gained by observing symmetries in the topological codes is to define representations of group actions on the logical output.
To illustrate this, group actions on the lattice express themselves as permutations of parity checks; in return, those permutations are directly associated with permutations of the syndrome and logicals. As investigated by Egorov et al. \cite{egorov2023end}, this information can be used effectively by extending the group averaging, to consider equivariance properties. This type of decoder is inspired by the concept of G-CNN, which are convolutional neural networks that can handle symmetry properties using groups. Cohen et al. proposed new types of layers that use symmetric groups for better weight sharing and theoretically higher performance in many tasks \cite{DBLP:journals/corr/CohenW16}. In their research, they used discrete groups on a square lattice, which perfectly overlaps with the standard lattice structure in quantum error correction. Using their G-CNN, they showed better performance than the traditional method, by exploiting symmetries in the data to improve error rates \cite{DBLP:journals/corr/CohenW16}. 

Unfortunately, the property of equivariance in the quantum case is harder to achieve than in the traditional sense, where functions act linearly as $\textbf{f}: V_{in} \rightarrow V_{out}$. In the case of Egorov et al., they encode equivariance properties using a matrix $M(\sigma)$, which directly depends on the syndrome \cite{egorov2023end}. Therefore, they argue that the standard theory on group equivariance no longer holds. The substitute for this will be discussed in the Methodology section.

In this work, we investigate the effects of more advanced decoder architectures, which have shown favorable performance in other domains \cite{bello2019attention, dosovitskiy2020image}.
Furthermore, we will extend the translational symmetry, inspired by the decoder architecture of Egorov et al. \cite{egorov2023end}, to the 3D case. We compare our approach with the non-trainable decoders BP-OSD \cite{panteleev2021degenerate} and SweepMatch\cite{Vasmer_2021}, both implemented in the PanQEC library \cite{panqec}.


\section{Methodology}\label{sec:methodology}
\subsection{Toric Code}
\begin{figure}[t]
    \centering
    \includegraphics[width=0.9\columnwidth]{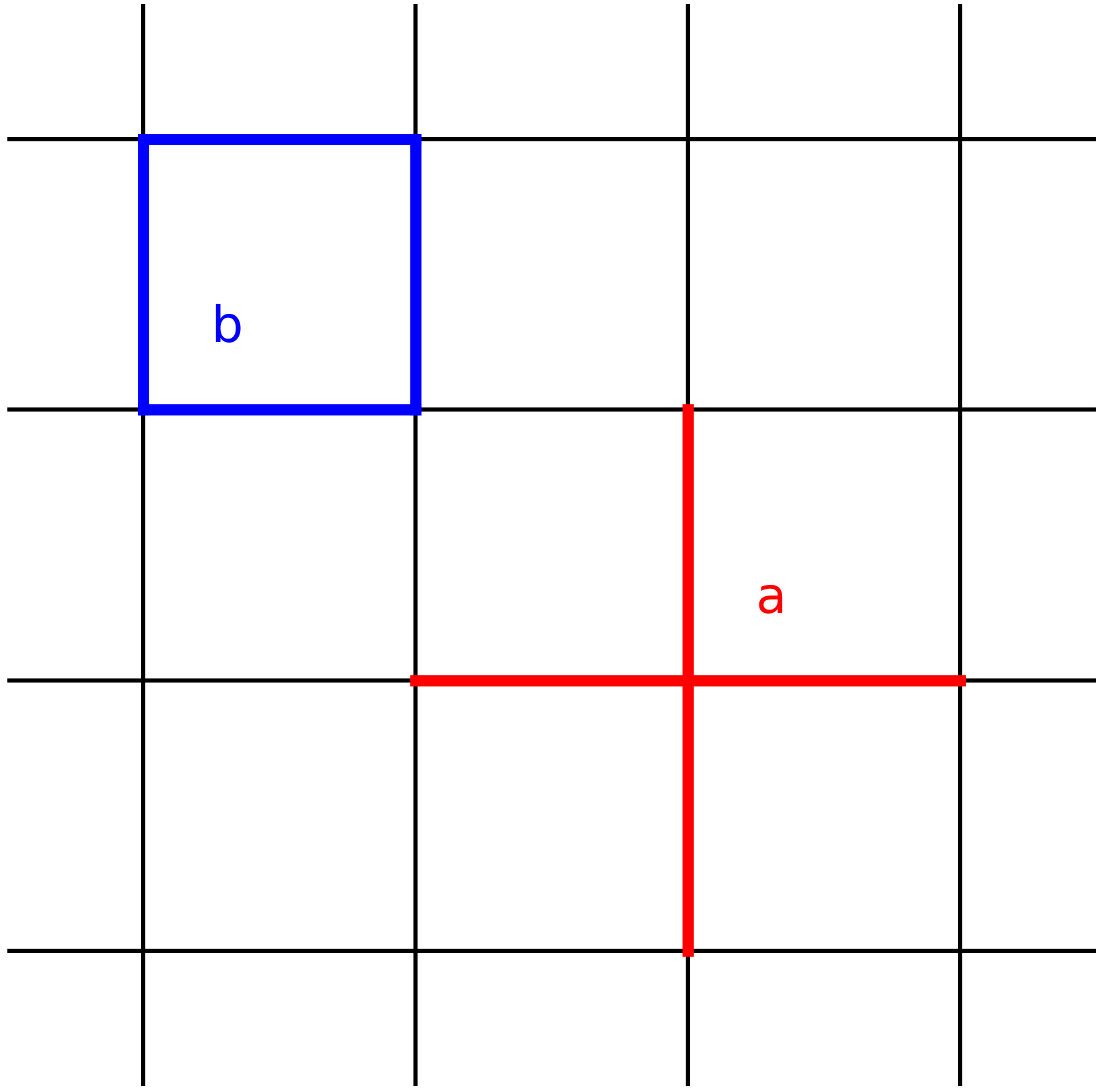}
    \caption{Lattice of 2D Toric Code, with X and Z Stabilizers \ref{eq:stabilizers}.}
    \label{fig:l2d}
\end{figure}

The toric code is one type of topological code frequently discussed in the literature due to its mathematical properties, ease of construction, and high locality of errors \cite{kitaev2003fault}. More concretely, the toric code is classified as a topological surface code, which means that the qubits are organized on a topological surface. The advantage of organizing the lattice on a torus is that we can increase the number of encoded qubits from 1 in the planar case to $d$, using a $d$-dimensional torus. Additionally, with the toric code, the exact location is not needed for errors to be corrected, which makes the code degenerate. In the two-dimensional case, the toric code can be constructed as a hypergraph product (HGP) of two classical repetition codes and, as such, is a hypergraph product code \cite{tillich2013quantum}. This construction gives us a square lattice as seen in Fig. \ref{fig:l2d}. 
Physical qubits are organized on the edges of the square lattice, encoding 4 logical qubits. These have a subspace of $2L^2$ physical qubits, with $L$ being the size of the classical codes.
Using this subspace, we can determine that any error-free code word lies within a 4-dimensional vector \cite{egorov2023end, bravyi1998quantum}. This error subspace is called `code word subspace` $C$ and can be thought of as a space that encodes the states of the logical qubits, depending on the states of the qubits in the lattice.
To measure the state of the qubits, we need stabilizers, which facilitate parity checks. On quantum codes, we have X and Z stabilizers, defined as products of bit- and phase-flips around a vertex. In the case of the 2D toric code, they are formed using the boundaries $\delta_i: C_i(K) \rightarrow C_{i-1}(K)$ and the co-boundaries $\delta^i: C^i(K) \rightarrow C^{i+1}(K)$ of the plaquettes $p$ and the vertices $v$, which tile the lattice (\ref{eq:stabilizers}). Here $C_i$ is the boundary map, with $C^i$ being the co-boundary map of $K$, a finite simplicial complex.
Stabilizers can also be thought of as errors that trivially affect the state of encoded qubits $S_i|\psi\rangle = |\psi\rangle \forall i$.  The X stabilizer in the 2D case is \textit{b} in Fig. \ref{fig:l2d} and the Z stabilizer is \textit{a}.

\begin{equation}\label{eq:stabilizers}
    S^Z_v = \prod_{n \in \delta^v} \sigma^x_n \quad, \quad S^X_p = \prod_{n \in \delta_p} \sigma^z_n
\end{equation}

\begin{figure}[h]
    \centering
    \includegraphics[width=\columnwidth]{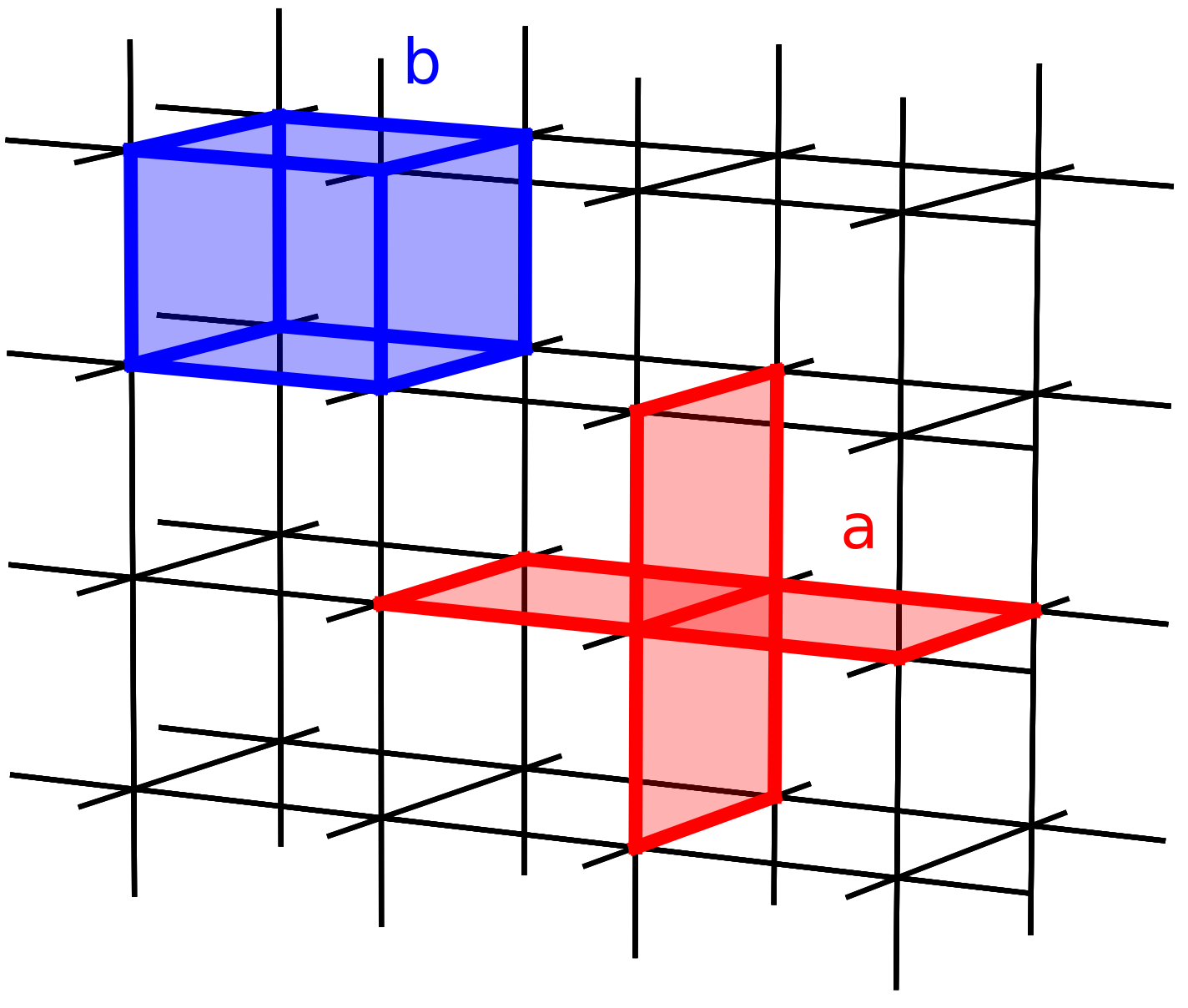}
    \caption{Lattice of 3D Toric Code, with X and Z Stabilizers.}
    \label{fig:l3d}
\end{figure}

These error-correcting codes can be represented using Tanner-graphs $\mathcal{T}$, therefore the construction of the toric code can be done using the hypergraph product,
denoted as $\boxtimes$ \cite{wang2020ak}.
A 3-dimensional code $\mathcal{T}_{t3d}$ is constructed by taking the HGP of three classical repetition codes $\mathcal{T}_{c}$, or the HGP of a 2D toric code $\mathcal{T}_{t2d}$ with another classical code (\ref{eq:hgp}). 
\begin{equation}\label{eq:hgp}
\begin{aligned}
   \mathcal{T}_{t3d} &= \mathcal{T}_c \boxtimes \mathcal{T}_c \boxtimes \mathcal{T}_c = \mathcal{T}_{t2d} \boxtimes \mathcal{T}_c\\
    \mathcal{T}_{t2d} &= \mathcal{T}_c \boxtimes \mathcal{T}_c 
\end{aligned}
\end{equation}
Similarly to the 2D case, the qubits again lie on the edges, while X stabilizers are now the 2-dimensional boundaries of the cubes that tessellate the lattice, and Z stabilizers are the 2-dimensional co-boundaries of edges, as seen in Fig. \ref{fig:l3d}. 

Note that in the 3D case, we now encode 6 logical qubits, with $3L^3$ physical qubits, and as such our code word subspace is 8-dimensional \cite{bravyi1998quantum}. The stabilizers in the 3-dimensional case can now be described as seen in (\ref{eq:stabilizers3d}), where $e$ are the edges and $c$ are cubes within the lattice. Furthermore, we can construct the logical operators $\mathbb{L}$ by taking stabilizers that form non-contractible loops around the torus manifold (\ref{eq:log}), across the unique axes (X, Y, Z), which can be denoted as abscissa, ordinate, and applicate ($ab$, $or$, $ap$), to avoid confusion with stabilizers. Therefore, each X-Z pair corresponds to one encoded qubit. Interestingly, in topological terms, the Betti number $b_1$ is directly related to the number of logical qubits encoded in the surface code. The most important result of the construction is the stabilizer matrix $H$, which allows us to model errors later.
\begin{equation}\label{eq:stabilizers3d}
    S^Z_e = \prod_{n \in \delta^e} \sigma^x_n \quad, \quad S^X_c = \prod_{n \in \delta_c} \sigma^z_n
\end{equation}
\begin{equation}\label{eq:log}
    \mathbb{L} =  (\bar{X}_{t}, \bar{Z}_{t}) \quad \forall \; t \in \{\text{abscissa, ordinate, applicate}\}
\end{equation}
We recall that the 3D Toric code consists of the HGP of three classical codes, and if we have symmetric lattice dimensions, we can construct the stabilizer matrix of the toric code using the Kronecker-product, as seen in Equation (\ref{eq:stab_matrix}). Here, $H_c(L)$ denotes the stabilizer matrix of the classical repetition code, with length L (\ref{eq:class}). $H_{t3d}(L)$ denotes the stabilizer matrix of the 3D toric code with size L. The final stabilizer matrix $H_{t3d}$ for $L=2$ is depicted in Fig. \ref{fig:stab_mtrx}.
\begin{equation}\label{eq:stab_matrix}
    \begin{aligned}
        c_x &= H_c(L) \otimes \mathbb{I} \otimes \mathbb{I} \\
        c_y &= \mathbb{I} \otimes  H_c(L) \otimes \mathbb{I} \\
        c_z &= \mathbb{I} \otimes \mathbb{I} \otimes H_c(L) \\
        H_{t3d}(L) &= \begin{bmatrix}
            & \text{\large{0}} & & c_x^T & c_y^T & c_z^T \\
            c_y & c_x & 0 & & &\\
            0 & c_z & c_y & & \text{\large{0}} \\
            c_z & 0 & c_x & & &\\
        \end{bmatrix}
    \end{aligned}
\end{equation}

\subsection{Error Model}\label{sec:err_model}
Once the lattice is constructed, errors can be modeled in different ways. Generally, an error model depends on the error rate $\textit{p}$, which determines whether an error affects a single qubit. Depolarizing noise is frequently used in the related literature, as it allows easy sampling and generation of data \cite{nielsen2001quantum}. Since errors in quantum systems can exhibit phase and bit flips, they are usually modeled using the Pauli group, from which we can use the X, Y, and Z operators to model errors, where the identity $\mathbb{I}$ does not represent an error. The equation to determine the probability of a specific error string $E$ with respect to the overall error rate $\textit{p}$ is described in (\ref{eq:error}).

\begin{equation}\label{eq:error}
\begin{aligned}
    p(E) &= \prod_{e \in \mathcal{E}} \pi(E_e)\\
    \pi(E) &= \begin{cases}
        1-p &  = \mathbb{I}\\
        p/3 & E\in \{X,Z,XZ\}.
    \end{cases}
\end{aligned}
\end{equation}

With $\mathcal{E}$ being the set of edges on the lattice, and [X, Z] being the Pauli operators that determine bit and phase-flip errors (\ref{eq:pauli}). Note that all errors $E$ can be mapped in the Pauli group representation. As such, we can ignore the Pauli $Y$ operator since it can be modeled as $XZ$, respectively.

Using the error model discussed, we can generate a large data set on the fly. Although easy to perform, this error modeling still represents the nature of errors in actual quantum circuits \cite{harris2020array}.
Note that an error string $E$ applied to the stabilizer matrix $H$ manifests itself as a syndrome. Due to the possibility that individual parts or an error cancel each other out, $E$ is not observable, but its syndrome $\sigma$ can be measured. This syndrome can be easily generated by applying noise to the stabilizer matrix ($H \times E$) seen in Fig. \ref{fig:syn_gen}.

\begin{figure}[h]
    \centering
    \includegraphics[width=\columnwidth]{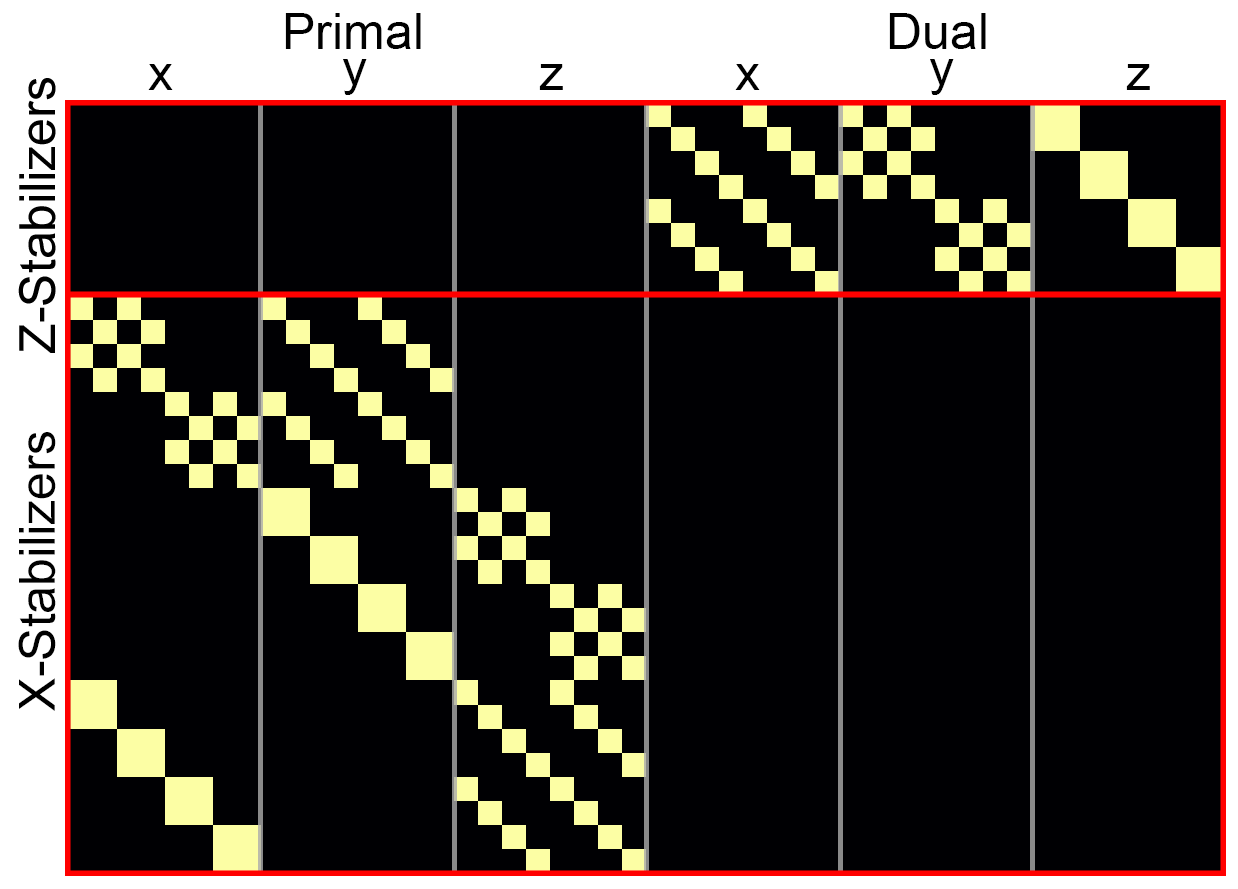}
    \caption{Stabilizer Matrix Toric 3D $L=2$}
    \label{fig:stab_mtrx}
\end{figure}

 Note that the syndrome can be decomposed into an X- and Z-part (corresponding to the stabilizers - see Fig. \ref{fig:stab_mtrx}). This syndrome can be represented as a matrix that forms the input to the decoder. The complete lattice of the toric code can be represented as a 3D matrix of size $L^3$, in which each value corresponds to a qubit on the lattice. Our syndrome, therefore, has a shape $4\times L^3$. 

The syndrome originally is a string of size $4\times L^3$, due to the representation of a primal and dual lattice. The primal lattice here is encoded in the first $3\times L^3$ components of the syndrome string, while the dual lattice is encoded in the last $L^3$ entries. In the 3D case, the syndrome can be split into four separate components (Fig. \ref{fig:syn_vox}). The most straightforward component is the syndrome on the dual-lattice, which can be measured using stabilizers that form surfaces in the manifold. However, the primal syndrome is measured using strings that form along one axis, which means that we now have three separate components to measure the complete syndrome. Each component can be reshaped into a tensor corresponding to the lattice shape.

\subsection{Evaluation}
To determine the performance of the various iterations of the machine learning-based decoder, we employ a simple evaluation measure, the logical accuracy. Recall that the objective of the error correction decoders is to reconstruct the most likely message given the syndrome $\sigma$. The network will then predict the logical components of the underlying error given the observed syndrome$\hat{\mathcal{L}}(\sigma)$. This allows us to determine whether the lattice is non-trivially affected by an error, and how it affects the encoded logical qubits. These logical components can be handled in multiple ways, as a multi-label classification problem, resulting in a result vector of size $\|\mathbb{L}\|$, where $\mathbb{L}$ are the logical operators of the code. Alternatively, they can also be one-hot encoded, which is the method used in our approach. This leads to an output vector of size $2^{||\mathbb{L}||}$. To determine the logical accuracy of this vector, categorical accuracy can be used as seen in (\ref{eq:acc}), where $N$ is the number of samples and $\mathbbm{1}$ is the indicator function.
\begin{equation}\label{eq:acc}
\begin{aligned}
    ACC &= \frac{1}{N}\sum^N_i \mathbbm{1}_{\mathcal{L}(E_i)}(\hat{\mathcal{L}}(\sigma_i))\\
    \mathbbm{1}_{\mathcal{L}(E_i)}(x) &= \begin{cases}
    1, & \text{if } x = \mathcal{L}(E_i)\\
    0, & \text{otherwise}
    \end{cases}
\end{aligned}
\end{equation}

In addition to the logical accuracy rate, we are also interested in the threshold capacity of the decoder $\bar{p}$. The threshold of a decoder can be thought of as an upper bound to the physical error rate, until which the decoder performs favorably. Once the physical error rate is above the threshold, the decoder cannot adequately handle the advantages of bigger lattice sizes anymore, and as such smaller lattices start outperforming again. This boundary is especially interesting to look at since it shows us the robustness of the decoder. Generally, each error-correcting code has a theoretical maximum threshold, $\bar{p} \sim0.11$ for the 3D toric code \cite{OHNO2004462}. Finding the threshold of the decoder is possible by plotting the logical accuracies of various lattice sizes and physical error rates and checking for the intersection of the resulting curves.

\subsection{Neural Decoder}
Our goal is to construct neural decoders to address the challenge of decoding the 3D Toric code. In our research, we take an iterative approach, developing multiple decoder architectures in which each successive version incorporates additional features. This allows us to easily compare different decoder versions and assess the impact of various components on performance.
For all proposed architectures, we use the AdamW optimizer, as it is efficient in the work of Egorov et al. \cite{egorov2023end}. AdamW also generalizes better than traditional Adam, due to its separate handling of weight decay \cite{loshchilov2017decoupled}. Furthermore, we initialize the convolutional layer weights using Kaiming-Normal initialization with Leaky-ReLU as nonlinearity. This approach was also used in \cite{egorov2023end} and has been shown to accelerate the convergence of deep CNN-type neural networks compared to standard Gaussian initialization \cite{he2015delving}. 

Furthermore, the loss function used with neural encoders is cross-entropy (CE), which is used to quantify the difference between the actual logical content of multiple errors $E$, indicated as $\mathcal{L}(E)$, and the predicted logical content of the error after observing the syndromes indicated as $\hat{\mathcal{L}}(\sigma)$. Note that errors $E$ can act trivially on the stabilizers and multiple errors can produce the same logical components. As such, the optimal way for the decoder is to predict whether any of the logical operators in $\mathbb{L}$ were affected by errors $E$. Furthermore, it is important to reiterate that errors $E$ are expressed in the lattices as syndromes $\sigma$. We are only able to observe the syndromes on the lattices due to the physical constraints of qubits. As discussed above, we can treat this problem as a multiclass classification or a multilabel classification; however, in our implementation, we only consider the multiclass problem (one-hot encoded labels), with a number of classes $\mathcal{C} = 2^{||\mathbb{L}||}$.

Due to the sampling method of noise strings, we expect a significant class imbalance. This imbalance depends on the lattice size and error rate per physical qubits, and as such is ever changing depending on the task. To counteract this in some way, we dynamically keep track of the logicals expressed in $\mathcal{L}(E)$, which we can then use as a weight $\omega$ for the loss function (\ref{eq:weight}). Here, $N$ is the number of samples in the batch and $N_{prev}$ is the number of samples seen before.
\begin{equation}\label{eq:weight}
    \omega = \frac{N_{prev} + N}
    {
    (\sum\mathcal{L}(E_{prev})+
    \sum\mathcal{L}(E))
    \mathcal{C}
    }
\end{equation}
From this, we define the weighted CE function as follows: (\ref{eq:bce}).
\begin{equation}\label{eq:bce}
    CE_\omega = \frac{1}{N}
    \sum^N_i
    \sum^{\mathcal{C}}_j
    \omega_j
    \mathcal{L}(E_i)_j \:
    \log\hat{\mathcal{L}}(\sigma_i)_j .
\end{equation}

For evaluating the decoders, we are using the logical accuracy of the model, which essentially describes how many of the errors could be mitigated properly using the decoder. This does not entail how good the encoder is at finding the exact error since we only predict the logical components, which simply say whether the error changed a logicals state or not. 

\subsubsection{Base Neural Decoder ($\phi_{GAP}$)}
To facilitate further expansions with equivariance properties and other improvements, we need a base structure that also serves as a baseline for experimentation. Generally, we need to construct a 3D CNN, to learn a map from the syndrome to the predicted logical class for the 3D Toric code. The cubic lattice in the code resembles a sequence of images or a video with four channels. The four channels originate from the parity check matrix (Fig. \ref{fig:stab_mtrx}), where each row corresponds to one channel in the final syndrome. Therefore, we have a wide variety of architectures to choose from. An especially powerful CNN- architecture originates from exploiting residual connections that essentially combine unprocessed elements with processed data, as proposed in the \textit{ResNet} paper \cite{he2015deep}. This type of expansion to traditional deep CNN architectures allows the network to optimize better and also gain more advantages from additional depth. Following the ideas behind the \textit{ResNet} architecture, \textit{WideResNet} was proposed, which instead of depth scaling, scales width. In theory, this makes networks easier to compute while still having comparable performance \cite{zagoruyko2016wide}. We adopt some elements of this architecture for our baseline model, as also done by Egorov et al. \cite{egorov2023end}. Throughout the network, we introduce non-linearity by using GELU activation, an alternative to traditional ReLu that showed good performance in various tasks \cite{hendrycks2016gaussian}.

For normalization within the network, we use standard batch normalization. Throughout the network, we employ simple 3D convolutions with cyclic padding, which follows from the fact that the lattice is embedded in a torus manifold having periodic boundaries. In addition to the classical convolutions, we added an attention mechanism to the initial and last convolutional layers. This has been shown to improve performance in \textit{ResNet}-style networks due to the addition of self-attention feature maps to convolution feature maps \cite{bello2019attention}. In our case, this did not drastically improve performance; rather it stabilized the convergence of the network; as such, we will not separately discuss the change of those layers.

For our baseline, we used this methodology and constructed a network inspired by \textit{WideResNet}.
To make the network more understandable, we showcase the construction of its building blocks here. Within the network, we use a number of wide residual blocks with the architecture seen in Fig. \ref{fig:wideblock}. These blocks can vary in depth, parameterized by $l$. In the base model, we set $l$ to 3 in all blocks.

In the proposed network architecture we use three \textit{WideResBlocks} with varying out-channels $\in \{128,64,64\}$. The kernel sizes have been set to 3 in all relevant operations.
The architecture of the base model can be seen in Fig. \ref{fig:wideresnet}.
Note that in the base case, we refrain from encoding the underlying symmetries in the lattice, therefore we use a traditional global average pooling layer (GAP) as the head of our CNN. 
This layer maps the output of the last convolutional layer to our output class vector ($\mathbb{R}^{(L)^3\times8\times(2)^3} \rightarrow \mathbb{R}^{64}$). To keep the maps readable, we notate n-dimensional matrices $\mathbb{R}^{a \times a \times a}$ as $\mathbb{R}^{(a)^3}$, which is different from one-dimensional matrices $\mathbb{R}^{a^3}$.
Using the index with the highest probability, we find the most likely encoded logical class. 
The concrete formula for the pooling can be seen in (\ref{eq:gap}), where $\phi: \mathbb{R}^{4\times(L)^3} \rightarrow \mathbb{R}^{(L)^3\times8\times(2)^3}$ denotes the CNN, and G is the product of three cyclical groups $\mathbb{Z}_{L}^{\times 3}$, essentially containing all the coordinates of potential points on the network. Note that the global average pooling layer averages channel-wise, and therefore the channel dimension is not shown in the equation, as it stays constant in size.
\begin{equation}\label{eq:gap}
    GAP = \frac{1}{|G|} \sum_{g\in G}\phi_g(\sigma)
\end{equation}

\subsubsection{Translational Equivariant Decoder ($\phi_{GAPT}$)}
Once the base decoder has been built and validated, more refined techniques are constructed with built-in equivariance depending on the lattice symmetries. The first property used in the adjusted decoder is \textit{translational}-equivariance. 
In traditional CNNs and tasks such as image recognition, translational equivariance is achieved by having periodic boundaries in convolutions. Meaning, a shift in the input will also shift the output accordingly. Using global average pooling as the head of the network, we achieve translational invariance for classification \cite{Goodfellow-et-al-2016}. However, this requires that linear actions on the input result in linear actions on the output. Unlike the consistent linear actions in CNNs, the output in our scenario can either stay the same or flip depending on the input, reflecting the variable nature of the logical errors. To clarify, logicals can remain unaffected by an error string but may become affected if the same error string is translated downward.
To formalize, we define the action of a group element $g$ on the stabilizers or errors, as seen in (\ref{eg:equivarence_}).
\begin{equation}\label{eg:equivarence_}
    S'_i = S_{gi}
\end{equation}
We follow \cite{egorov2023end} and extend 2d lattice to 3d. The translational symmetries within the 3D lattice can be split into three distinct parts: abscissa-translation (X-axis), ordinate-translation (Y-axis), \ and applicate-translation (Z-axis). The translations for the errors $X$ and $Z$ are formulated as in (\ref{eq:trans_sym}), where $T$ is the set translations $T$ = {abscissa, ordinate, applicate}.  Recall that CNNs, by nature, are translationally invariant due to the pooling layers used \cite{Goodfellow-et-al-2016}. However, this does not entail translational equivariance as discussed above. Therefore, for our adaptation, we extend the idea of the global average pooling layer using a check matrix $M$, which entails conditional checks to determine whether a probability computed by the neural network $\phi$ should be swapped ($p \rightarrow 1-p$).

\begin{equation}\label{eq:trans_sym}
    \bar{Z'_t}= \bar{Z_t}\prod_{n\in \delta^e}S^Z_n, \quad
    \bar{X'_t}= \bar{X_t}\prod_{n\in \delta_c}S^X_n
    \quad \forall \quad t \in T
\end{equation}

\begin{figure}[t]
    \centering
    \includegraphics[width=\columnwidth]{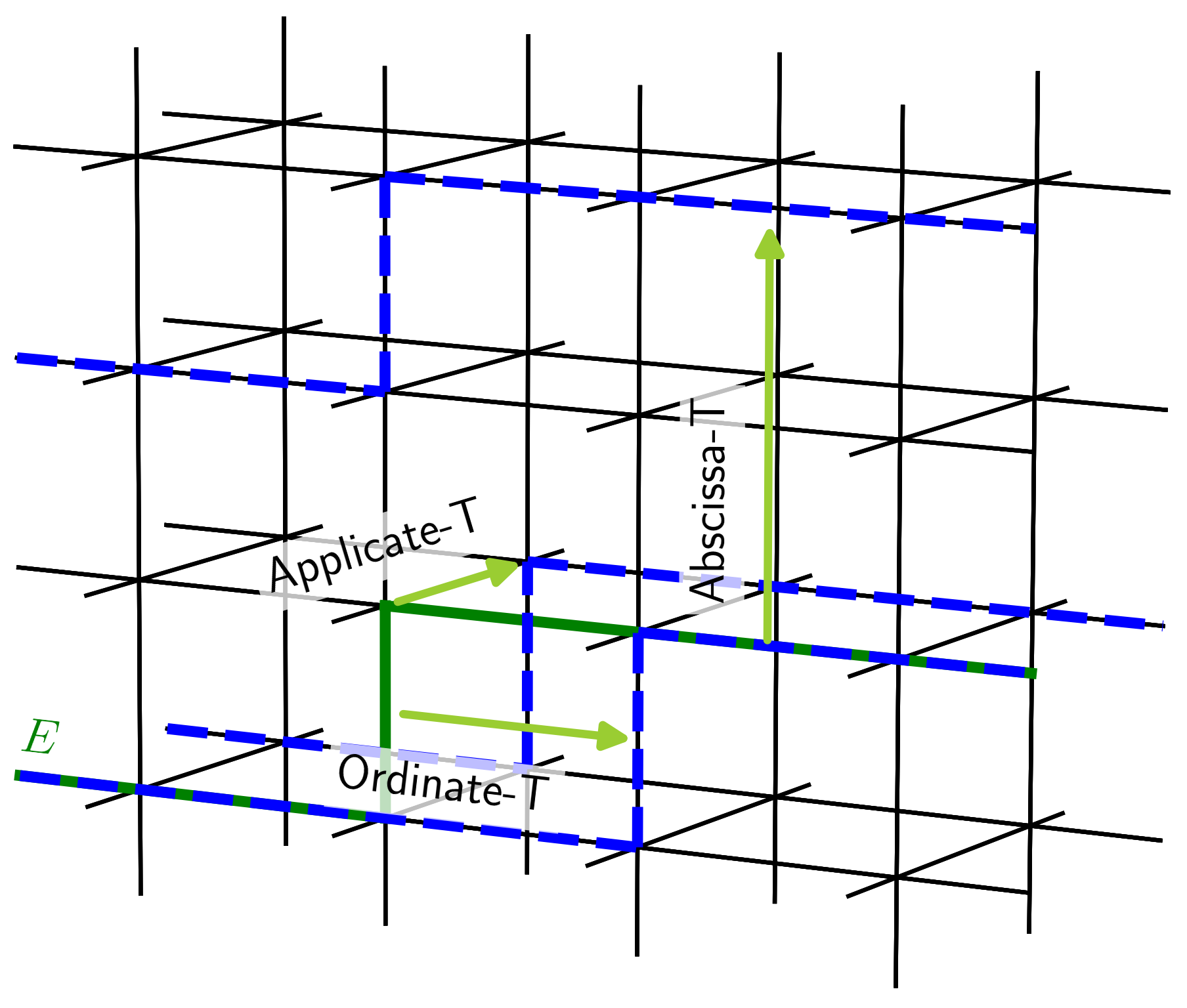}
    \caption{Translational Symmetries on the 3D Lattice}
    \label{fig:trans_sym}
\end{figure}

To visualize the translation symmetries in the lattice, see Fig. \ref{fig:trans_sym}, where an error $E$ is translated along the X, Y, and Z axes to obtain different $E'$. 
Concretely we formulate the adapted pooling in this example by multiplying the output of the CNN by our logical probability matrix, as proposed by Egorov et al. \cite{egorov2023end}. The translationally equivariant pooling layer is then formulated as seen in (\ref{eq:gapt}).

\begin{equation}\label{eq:gapt}
    GAP_T =  \frac{1}{|G|} \sum_{g\in G}M_g(\sigma)\phi_g(\sigma)
\end{equation}

The logical probability matrix $M$, depending on a transformation $g$ is constructed as seen in Equation (\ref{eq:m_const}). Note that $a$ is simply an index to the codespace we have on the lattice (size $2^d$, with $d$ being the dimensionality of the code).
$P_g$ is a permutation operator that is applied to the logical probability tensor $t = \phi(\sigma)$; however, for the sake of translational equivariance, it can be left out, since it acts as the identity. This permutation will be relevant in rotational equivariance. In equation (\ref{eq:delta_g}) $\alpha = \delta^e$, and $\beta = \delta_c$, concretely they are aliases for the co-boundaries of edges and boundaries of cubes.

\begin{equation}\label{eq:m_const}
    M_g(\sigma) = P_g^{-1}\prod^{2^d}_{a=1}R^g_a(\sigma)
\end{equation}
\begin{equation}\label{eq:r}
    R_a^g = 
    \begin{cases}
    \mathbb{I} &\mbox{ if }  \triangle_g(g^{-1}\sigma)_a \mbox{  mod } 2 = 0 \\
    t_{...\gamma_a...} \rightarrow t_{...1-\gamma_a...} & \mbox{ otherwise }
        
    \end{cases}
\end{equation}
\begin{equation}\label{eq:delta_g}
    \triangle_g(\sigma)_a = \sum_{v\in\alpha^g_a}\sigma^Z_v + \sum_{p\in\beta^g_a}\sigma^X_p
\end{equation}

\subsubsection{Transformer Stacked Decoder ($\phi_{ViT}+\phi_{GAPT}$)}
In recent years, transformer network architectures have gained significant traction due to their high performance across various domains. Although CNNs were once state of the art for vision-related tasks, they have gradually been overshadowed by vision transformers (ViT) \cite{dosovitskiy2020image}, which offer greater flexibility and better generalization in a range of tasks \cite{bai2021transformers}.

However, due to the complexity of decoding the toric code in 3D, using a pure transformer architecture did not yield acceptable results. This is similar to the base decoder ($GAP$), which struggled without additional knowledge of symmetries. In particular, the ViT exhibited a much higher initial accuracy and a more uniform accuracy across all logical errors compared to the base CNN, which was better at extracting simple logical errors affecting only one logical.

Unlike other studies that typically focus on pure CNN or pure transformer architectures, we propose a novel approach combining both. To take advantage of the strengths of both methods, we combine the superior local- 
and symmetry-knowledge of the CNN with the global knowledge of the ViT. This results in a stacked decoder ($ViT+GAP_T$), or the final prediction is calculated as: $\frac{1}{2}(\phi_{ViT}(\sigma) + \phi_{GAP_T}(\sigma))$. This combination approach demonstrates the potential benefits of integrating CNNs and ViTs for complex decoding tasks, offering an alternative to predominantly \textit{ pure} architectures commonly used in other studies.

\begin{figure*}[t]
    \centering
    \includegraphics[width=0.7\textwidth]{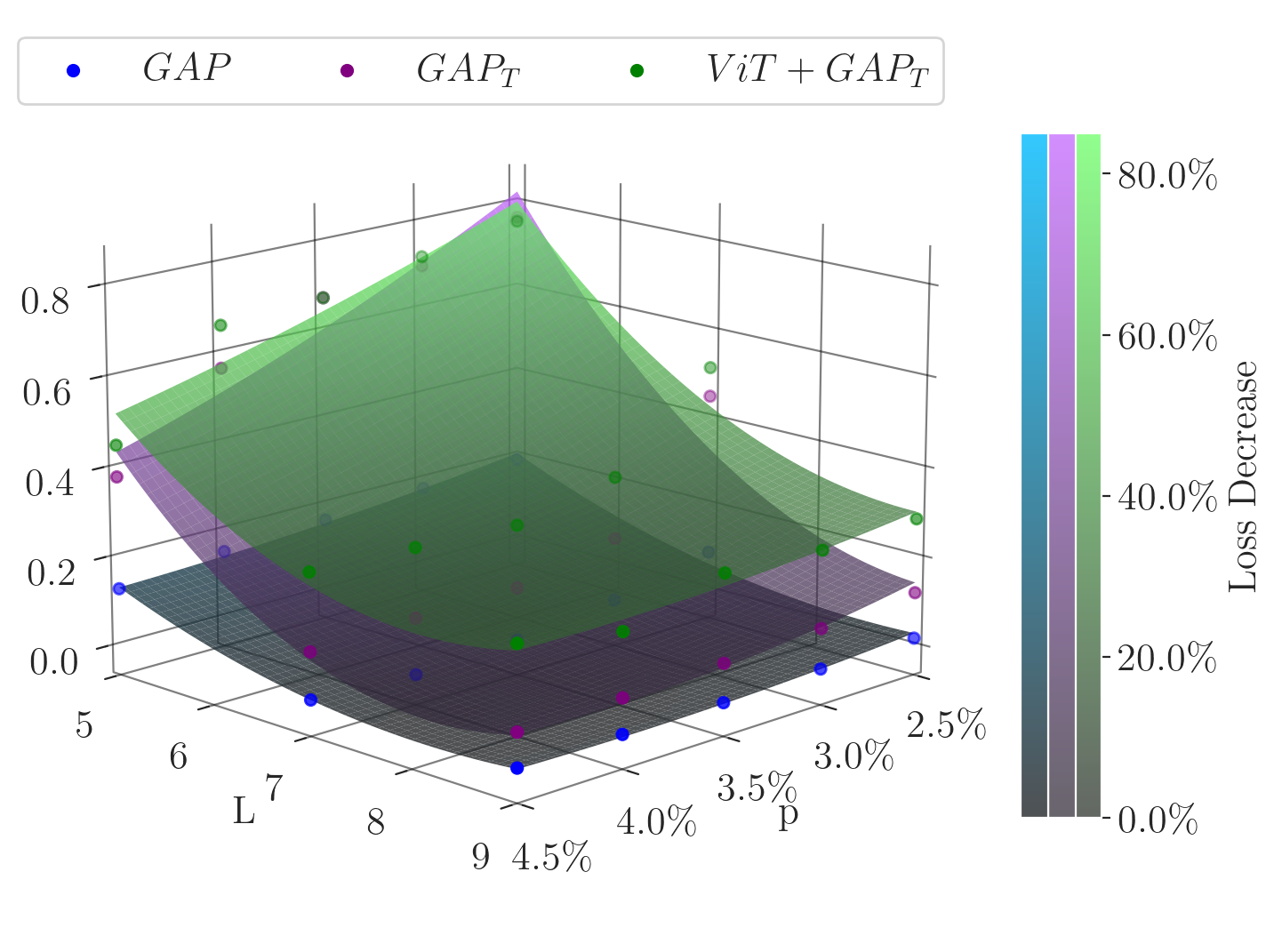}
    \caption{Trainability Surface of Decoders}
    \label{fig:trainability}
\end{figure*}
\subsection{Experimental Setup}
To validate the proposed networks, we will train each iteration on $10^6$ sampled syndromes with their logical content class as a label. Those data points are dynamically generated using our error model described above. For evaluation, we use the logical accuracy described in Equation (\ref{eq:acc}), and the loss function in Equation (\ref{eq:bce}). The network training itself is performed by generating new batches on the fly until we obtain a total number of $10^6$ sample data points. We train with a batch size of $512$ and have a maximum learning rate of $0.1$ for the AdamW optimizer, which is scheduled using OneCycle scheduling \cite{smith2018superconvergence}. Each decoder will be trained with various physical error rates $p_{train}$, using a binary search-like order to find the highest $p_{train}$, which can still maintain an accuracy of $>0.5$.

For evaluation, we take the BP-OSD decoder used by \cite{Quintavalle_2021, roffe2020decoding, panteleev2021degenerate} and the SweepMatch decoder\cite{Vasmer_2021} as our baselines. The evaluation experiments will vary in the physical error rate, ranging between $p \in {0.021...0.048}$. This range has been chosen since the 3D toric code BP-OSD has reached a threshold of around $0.029$ \cite{Quintavalle_2021}. We also vary the size of the lattices $L \in {5,7,9}$. Note that we do not retrain the models for different error rates and lattice sizes and as such test the generalizability of the networks in varying environments. Finally, once all the experiments have been completed, we can estimate the thresholds of the decoders and compare their performance with the baselines.

All code for the network architectures and experiments is available under: \url{https://github.com/oliverweissl/NeuralDecoderToric3D}.


\section{Results}\label{sec:results}

\subsection{Trainability}
Following the experiments, we are interested in multiple different results.
While accuracy tells us whether the decoder is decoding correctly, loss is an indicator of how incorrect the predictions are. By using the loss, we adapt the weights in the network, to ideally have a smaller loss in the next iteration. Generally, when the loss is decreasing, we acknowledge this as "training" of the network. However, when the loss stays constant, either the network is faulty or it is an indicator that it is not powerful enough to capture the correlations in the data. 
As such loss can be seen as a proxy to \textit{trainability} of a certain decoder with a certain task. Higher trainability would be indicated by a more drastic decrease in the loss over training iterations.
The trainability surface in Fig. \ref{fig:trainability} (alternatively, Fig. \ref{fig:train2d}) shows this notion, which is essentially the relative decrease in loss over all training iterations for the proposed decoder architectures. The higher the value, the better the model captures the problem's characteristics. This figure allows us to visualize the effects of improvements on the neural decoder. Interestingly, our more advanced architectures are "lifting" the surface indicating that architecture changes can drastically improve trainability.

\subsection{Decoder Performances}
Another important aspect is the performance of our decoder architectures and a comparison between them. We do so not only by looking at the final accuracy they achieved but also the highest error rate on which they can be trained while still having sufficient accuracy ($>0.5$).
In addition to the decoder-specific metrics, we are also interested in the threshold $\bar{p}$ because it shows how robust the decoder is overall. The threshold can be thought of as an upper bound of the error rate on which the decoder can perform favorably. Ideally in topological codes, we want to add more qubits to ensure better encoding performance for the logical qubits; however, the threshold dictates until which error rate we can exploit the benefits of increased lattice sizes.
\begin{table}[h]
\caption{Decoder Performances}
\resizebox{\columnwidth}{!}{%
\begin{tabular}{cllllll}
\hline
\multicolumn{1}{l}{\textbf{Decoder}} &
\textbf{L} & \textit{p:} & \textbf{2.1\%} & \textbf{3.0\%} & \textbf{3.9\%} & \textbf{4.8\%} \\ \hline \hline
                             & 5 & $p=0.1\%$: 0.783 & 0.004          & \textless 0.001 &       &        \\
                             & 7 & $p=0.1\%$: 0.499 & \textless 0.001 &                 &       &        \\
\multirow{-3}{*}{SweepMatch\cite{Vasmer_2021}} & 9 & $p=0.1\%$: 0.231 & \textless 0.001 &                 &       &        \\ \hline
                             & 5 &                & 0.975           & 0.913           & 0.801 & 0.6709 \\
                             & 7 &                & 0.983           & 0.905           & 0.734 & 0.5047 \\
\multirow{-3}{*}{BP-OSD\cite{panteleev2021degenerate}}     & 9 &                & 0.987           & 0.900           & 0.645 & 0.3658 \\ \hline \hline
                             & 5 & $p_{train} = 4\%$ &  0.332 &  0.288 & 0.240 & 0.198 \\
                             & 7 & $p_{train} = 2.5\%$ & 0.198 & 0.138 & 0.101 & 0.077 \\
\multirow{-3}{*}{$GAP$}        & 9 & $p_{train} = 1.25\%$ & 0.135 & 0.085 & 0.068 & 0.059 \\ \hline
                             & 5 & $p_{train} = 4\%$ & 0.576 & 0.600 & 0.524 & 0.420 \\
                             & 7 & $p_{train} = 2.5\%$ & 0.466 & 0.333 & 0.150 & 0.065 \\
\multirow{-3}{*}{$GAP_T$}    & 9 & $p_{train} = 1.25\%$ & 0.161 & 0.060 & 0.045 & 0.039 \\ \hline

                             & 5 & $p_{train} = 4\%$ & 0.872 & 0.785 & 0.686 & 0.577  \\
                             & 7 & $p_{train} = 2.5\%$ & 0.624 & 0.480 & 0.347 & 0.233  \\
\multirow{-3}{*}{$ViT+GAP_T$} & 9 & $p_{train} = 1.25\%$ & 0.298 &  0.193 & 0.127 & 0.089 \\ \hline
\end{tabular}}
\label{tab:perf}
\end{table}
We benchmark the non-trainable decoders SweepMatch\cite{Vasmer_2021} and BP-OSD \cite{panteleev2021degenerate} on $10^5$ samples, due to their computationally expensive nature. As seen in Table \ref{tab:perf}, the SweepMatch decoder does not perform well on combined Phase- and Bit-Flip noise, and does not achieve meaningful results with error probabilities $p > 0.001$. The BP-OSD decoder, on the other hand, achieves favorable performance in error rates ranging between $0.021-0.048$. In addition to the non-trainable decoders, we show the performance of the base decoder ($GAP$), the translationally equivalent decoder ($GAP_T$), and the final vision transformer ensemble ($ViT+GAP_T$). For trainable decoders, $p_{train}$ is given in Table \ref{tab:perf}, with each decoder evaluated on $10^6$ samples.

\begin{figure}[h]
    \centering
    \includegraphics[width=\columnwidth]{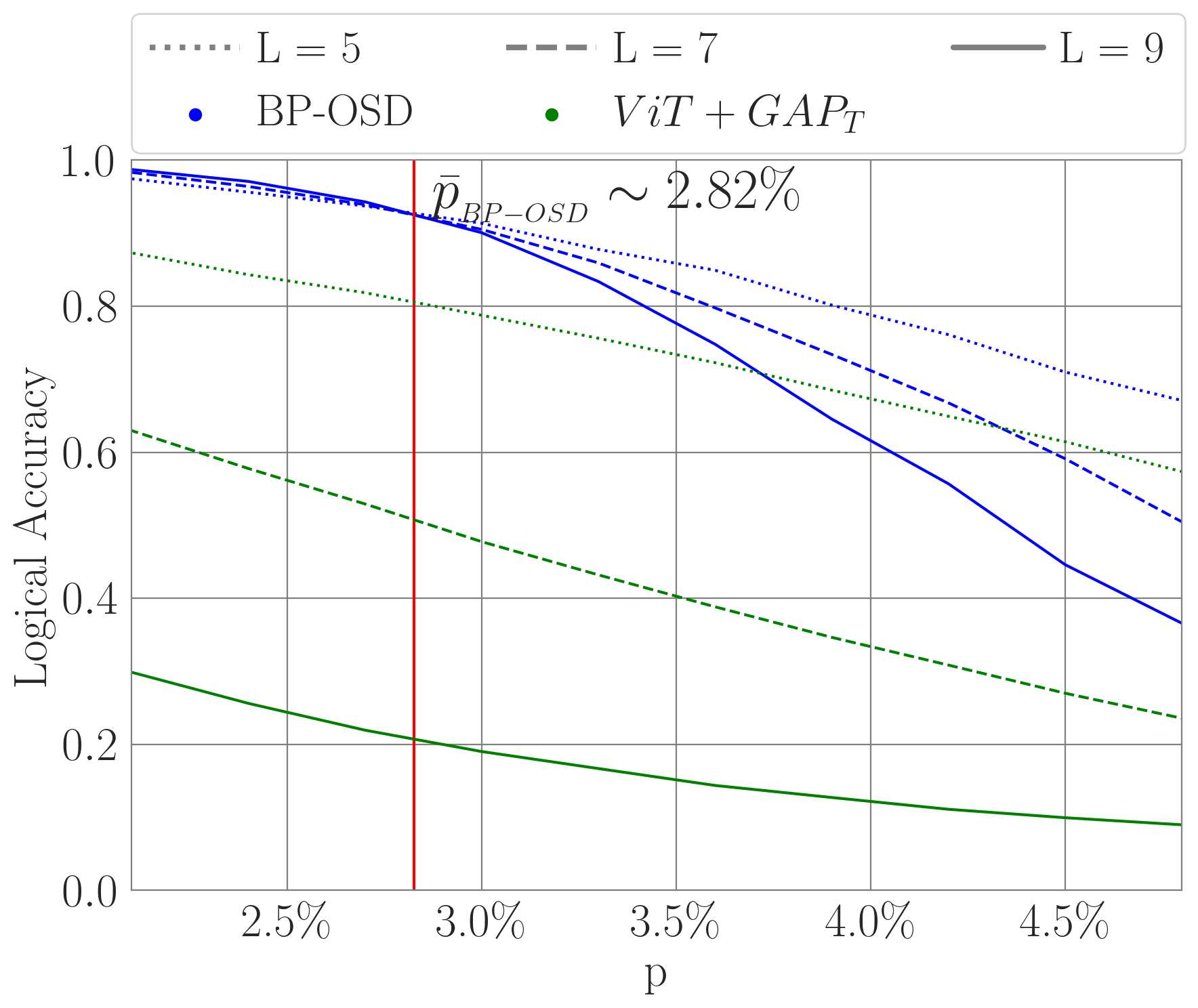}
    \caption{Performance of Decoders}
    \label{fig:thresh_plot}
\end{figure}

For each lattice size $L$, we plot the performance of the $ViT+GAP_T$ and $BP-OSD$ decoders in Fig. \ref{fig:thresh_plot}.
The threshold itself can be determined by inspecting the performance curves, once the bigger lattice sizes start underperforming smaller lattices, we expect the threshold $\bar{p}$. For the BP-OSD decoder, the threshold in our case was $\bar{p}\sim0.028$ as seen in Fig. \ref{fig:thresh_plot}.

\begin{figure}[h]
    \centering
    \includegraphics[width=\columnwidth]{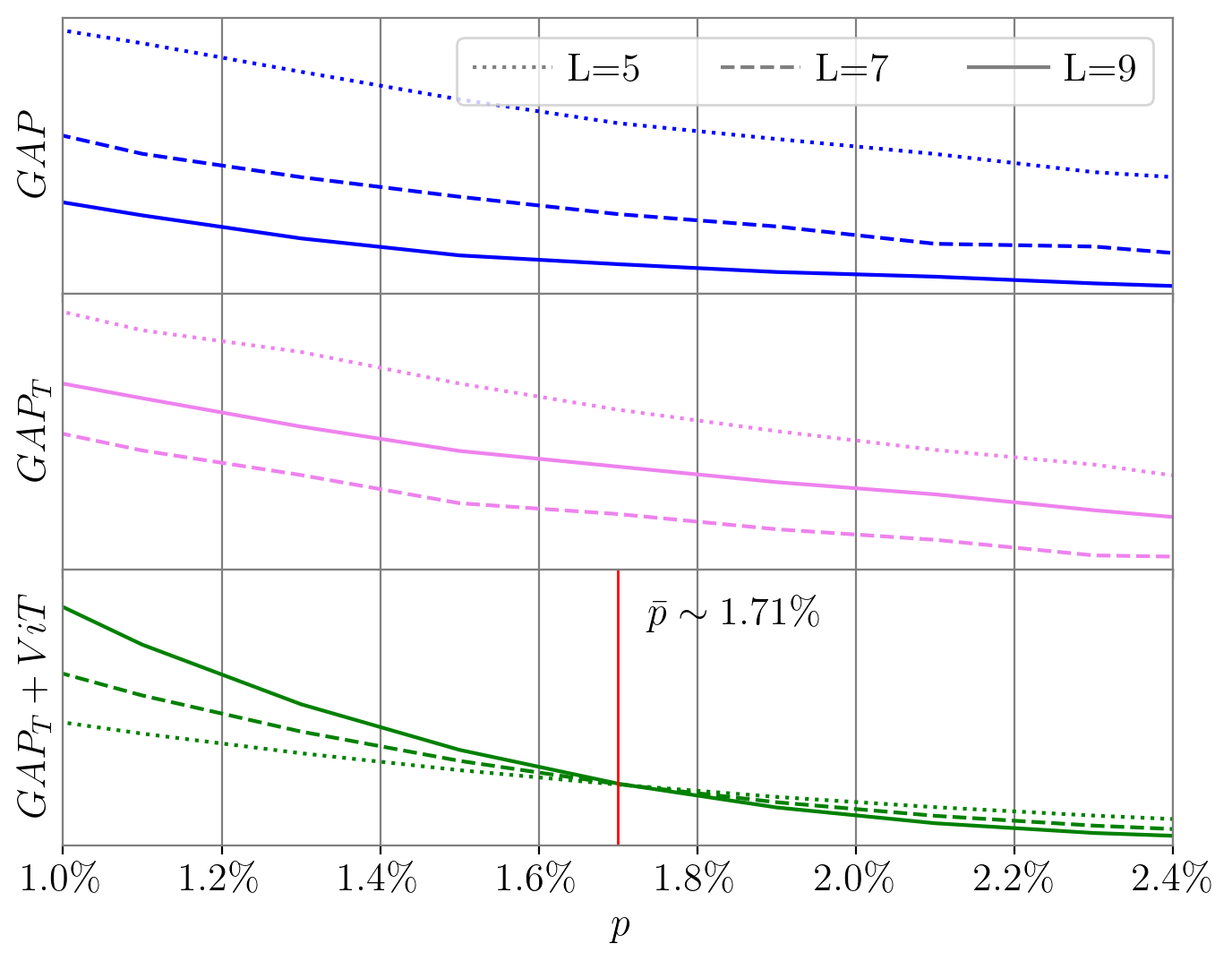}
    \caption{Threshold discovery}
    \label{fig:thresh_analysis}
\end{figure}

However, with $ViT+GAP_T$, we only plot the best decoder for each $L$. Those decoders were trained on different error rates $p$ and as such we cannot use those curves to determine the threshold. To evaluate the threshold of the decoder, we plot a trained decoder on a \textbf{fixed} $p$, with varying values for $L$.
In Fig. \ref{fig:thresh_analysis} we can see the performance curves of all iterations of the trainable decoders, with $p = 0.0125$. Note that for $ViT+GAP_T$, the vision transformer is not flexible with respect to L, due to the rigid nature of the positional embeddings and, as such, is left out of the evaluation. We find that the thresholds for $GAP$ and $GAP_T$ are $\bar{p} < 0.001$, however, using $ViT+GAP_T$ we achieve $\bar{p} \sim 0.0171$.

\subsection{Runtime Analysis}
\begin{figure}[H]
    \centering
    \includegraphics[width=\columnwidth]{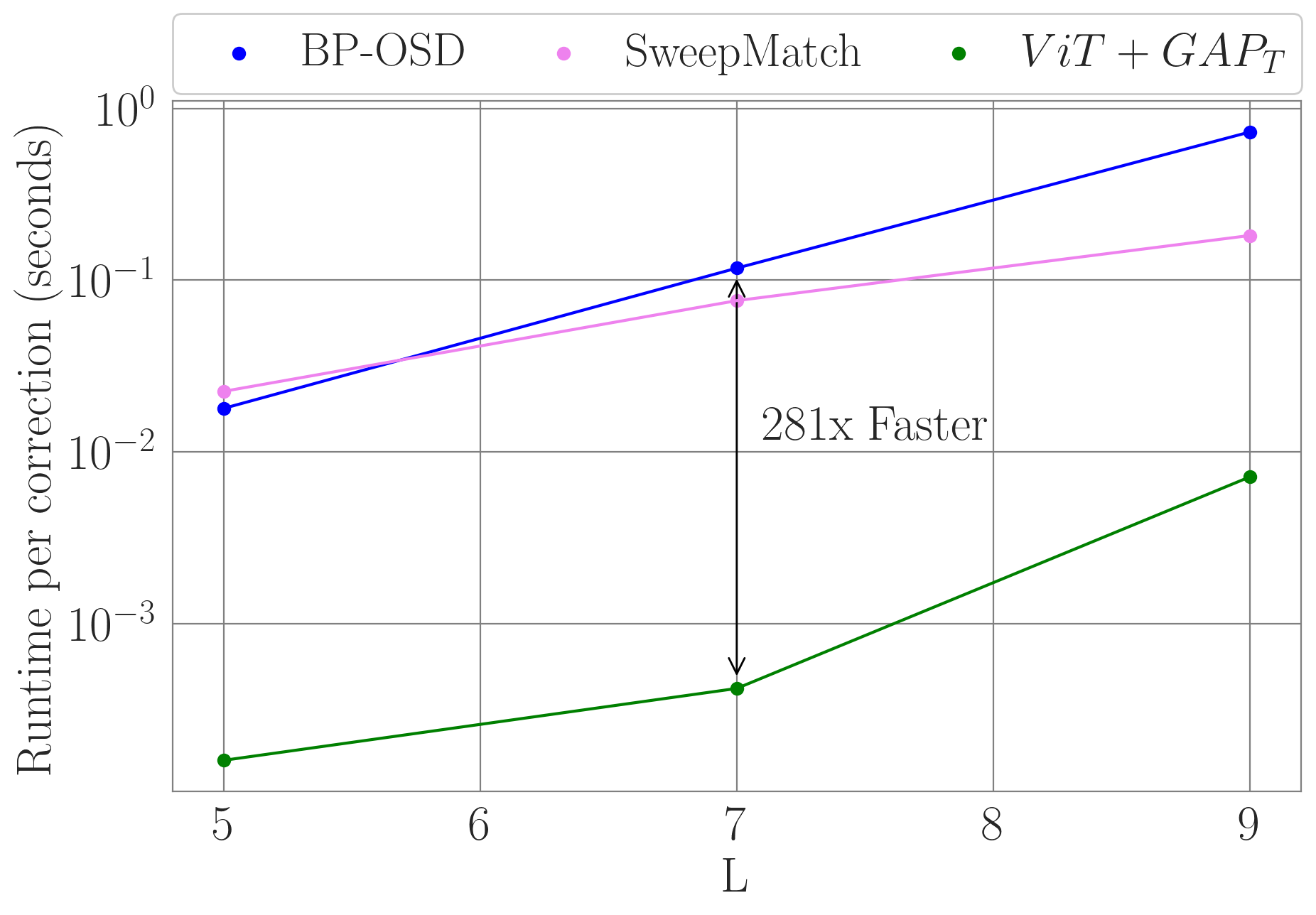}
    \caption{Runtime of Decoders}
    \label{fig:runt_rel}
\end{figure}

While runtime is generally not a concern in most related research, error-correcting algorithms in production do have to deal with the so-called "reliability/data rate" trade-off \cite{tse2005fundamentals}. Although theoretically, we could have a close-to-perfect reconstruction of errors, these reconstructions must be made promptly to have relevance. In traditional communication technology, we would not benefit from WIFI taking tens of times longer for an additional inch toward reliability. The same trade-off must be considered in QEC once it is production-ready. Although the neural decoder did not perform better in terms of accuracy, the difference in run time is dramatic, as shown in Fig. \ref{fig:runt_rel}. Considering that the runtime is heavily dependent on hardware, this figure illustrates the difference in performance potential, not absolute speedup. Additionally, SweepMatch and BP-OSD benefit from multi-shot decoding (amplifying the already big performance difference), while the neural decoder generally is one-shot.


\section{Discussion}

\label{sec:discprovenown}
As the field of quantum error correction (QEC) gains importance, new, novel, and more powerful methods are being developed. One such novel approach was first proposed by Egorov et al. \cite{egorov2023end}, exploiting the symmetries of square lattices. In this research, we expand on this approach and generalize it to cubic lattices. 
Benchmarking our proposed decoder architectures, we found that including translational equivariant pooling is an essential building block to make neural decoders (CNN architectures) trainable. As seen in Table \ref{tab:perf}, the neural decoder using a standard global-average-pooling $GAP$ (\ref{eq:gap}) performs rather poorly in the given task. But without changing the network architecture and only switching the pooling approach to $GAP_T$ (\ref{eq:gapt}), performance improved significantly, especially for smaller lattice sizes. 

Although we saw a drastic increase in performance, the comparison decoder BP-OSD \cite{panteleev2021degenerate}, still outperformed by a high margin. To reduce the gap, we added a vision transformer \cite{dosovitskiy2020image} to the decoder, resulting in an ensemble decoder $ViT+GAP_T$. Although not as powerful as BP-OSD, we managed to increase the performance of trainable decoders by up to $250\%$ compared to the implementation of $GAP$.

The threshold of BP-OSD was $\bar{p}\sim0.028$, which is much higher than the threshold of our ensemble method $ViT+GAP_T$, where $\bar{p} \sim 0.0171$. However, an interesting observation in the threshold discovery is that without $GAP_T$ and the $ViT$ ensemble, the expected $\bar{p} < 0.001$, which shows that novel additions to the standard CNN architectures are needed to make QEC with higher dimensional codes feasible (see Fig. \ref{fig:thresh_analysis}). Another interesting observation from these results is that, using ensembles, the decoder can better exploit the strengths of different architectures. Note that in Fig. \ref{fig:thresh_analysis}, $GAP_T$ and $ViT+GAP_T$ only rely on the CNN part with its $GAP_T$ pooling. This is due to the positional embeddings in the vision transformer that do not allow changes in $L$. Although both decoders theoretically use the same components in this scenario, $\bar{p}$ for the ensemble is much higher than for $GAP_T$ alone. This suggests that by using ensembles the separate parts learn to classify different types of error. Note that CNNs generally excel at extracting local information, whereas transformer architectures are better at using global information. This difference in strengths manifested itself in a drastic increase in error-correcting performance.

Although our decoders performed worse in terms of accuracy (compared to BP-OSD), we still had a significant advantage over BP-OSD in runtime.
BP-OSD benefits from multiple correction cycles and is very computationally heavy, making it slow on the systems we tested it on (Table \ref{tab:specs}). The neural decoders require significant training effort initially, but the correction attempts were faster by orders of magnitude (compared to BP-OSD and SweepMatch), as shown in Fig. \ref{fig:runt_rel}. This trade-off between speed and accuracy needs to be considered once QEC methods are used in production environments \cite{tse2005fundamentals}. Neural decoders offer a flexible and fast alternative to non-trainable decoders, which are more exact but also slower overall.

The chosen decoder architecture has proven its efficiency in error correction and has gotten favorable results\cite{egorov2023end}, but we must consider that the 3D case is more computationally complex compared to the 2D case. This suggests that increasing the network size might improve the accuracy of the decoders to some point. However, the most reliable way to increase performance is to adapt the architecture of singular components to better capture the complexities of the problem. As seen with the increase in performance and trainability (Fig. \ref{fig:trainability}) using translational equivariant pooling, we can theoretically exploit more equivariance properties in the cubic lattice. One major candidate for this would be rotation; as with cyclic group rotation, we could better generalize the effects of error paths. We did not implement a component to take advantage of these symmetries; however, future research could explore this path further.


\section{Conclusion}
\label{sec:conclusion}
Due to the fragility of quantum information, quantum error correction strategies are required to enable quantum computing systems to scale \cite{gambetta2017building, GYONGYOSI201951}.

Many powerful error-correcting algorithms and strategies have been developed; however, most do not consider the inherent symmetries of the lattices in error-correcting codes \cite{PhysRevA.100.012312, Quintavalle_2021, panteleev2021degenerate}. Furthermore, these methods usually rely on many error-correcting cycles, making them slow to execute.

Neural decoders, on the other hand, provide a lightweight (trained) map $f: \sigma \rightarrow \hat{\mathcal{L}}(\sigma)$, which takes significant initial computing power to establish, but once computed it is very fast to produce results. These neural decoders usually suffer from low accuracy and a high data requirement due to the complex nature of syndromes on the error code lattices. By embedding error-correcting codes in topological constructs, we can exploit inherent symmetries, which have been shown to significantly improve the performance of decoders on the 2D toric code\cite{egorov2023end}.

In this paper, we explore how a CNN-based maximum likelihood decoder can be adapted to encode equivariance properties of the cubic lattice, embedded in a 3D toric code.
Our first research sub-question regards how the equivariance properties of the maximum likelihood decoder can be described for graphs embedded in 3-dimensional manifolds.
In this research, we focus on the translational equivariance property, which is formulated in Equation (\ref{eq:trans_sym}). The effects of translations on the error string can be seen in Fig. \ref{fig:trans_sym}.

Following the definition of the translational equivariance, we can answer the second research sub-question "How can those equivariance constraints be embedded in a neural decoder architecture?".
Building on the extension of group averaging described in \cite{egorov2023end}, we expand the global average pooling layer into a translational equivariant pooling layer $GAP_T$ (\ref{eq:gapt}). Our twist to $GAP$ is the condition matrix $M$, first described in \cite{egorov2023end}, which we construct for the three-dimensional lattice as shown in Equations (\ref{eq:m_const}, \ref{eq:r}, \ref{eq:delta_g}).

After successfully constructing an equivariant pooling layer, we aim to benchmark the performance of this decoder with respect to other decoders available on the 3D Toric code. In our experiments, we specifically chose SweepMatch \cite{PhysRevA.100.012312} and BP-OSD  \cite{panteleev2021degenerate}, due to their implementation in the PanQEC library \cite{panqec} and their suitability for the 3D toric code. From Table. \ref{tab:perf} we can see that BP-OSD has better performance than all neural decoders explored in this paper; however, it is significantly slower in run-time (Fig. \ref{fig:runt_rel}). This showcases an interesting trade-off in accuracy and computational time, widely considered in traditional error correction research but not in the focus of current QEC papers \cite{tse2005fundamentals}.

Although not trumping BP-OSD in terms of error threshold and accuracy, we showed that using trainable decoders in QEC (specifically on the 3D Toric code) is a viable option, given that their architecture is adapted to reflect the complexities of the task. As seen in Tab. \ref{tab:perf} and Fig. \ref{fig:thresh_analysis}, using a pure CNN decoder with traditional $GAP$ pooling is unattainable due to poor training performance. However, when introducing equivariance properties such as the $GAP_T$ pooling (\ref{eq:gapt}), trainability (Fig. \ref{fig:trainability}) and performance rise dramatically. However, because the 3D lattice has an inherited mismatch in the expression of X and Z errors in the lattice \ref{sec:err_model}, translational equivariance was insufficient to show a sustainable error threshold greater than 0.001. Using a vision transformer \cite{dosovitskiy2020image} we expanded the decoder, exploiting CNN`s local knowledge and Transformer`s global knowledge. This ensemble approach $ViT+GAP_T$ produced a respectable error threshold of $\sim 1.71\%$, further increasing performance significantly.

In this paper, we showed that trainable decoders are viable options for QEC. Although we did not outperform BP-OSD`s accuracy, the correction speed with our neural decoders was superior. Additionally, we established that intelligent component design of neural decoders can significantly improve task performance, paving the way for future research to explore the potential of adding rotational equivariance. On another note, we want to stress that the choice of architecture greatly impacts performance. Although a pure transformer-based approach was unable to learn the task, similar to the $GAP$ approach, combining CNN with ViT showed a great jump in performance. This jump hints at the importance of both local and global information. Future research should incorporate these different strengths and explore how positional embedding in the ViT can be generalized so that varying lattice sizes can still be used with a single decoder instance.


\bibliography{main}
\clearpage

\appendix
\begin{appendices}

\section{Methodology}
\subsection{Toric Code}
The construction of the parity-check matrix for a classical repetition code of size $L$ can be formulated as seen in (\ref{eq:class}). Note that here $\delta$ is the Kronecker delta function, with $i, j \in \{0,1,…,L-1\}$.
\begin{equation}\label{eq:class}
    H_c(L)_{i,j} = \delta_{i,j} + \delta_{i,(j+1)mod\; L}
\end{equation}

\subsection{Error Model}\label{sec:app_err}
\begin{figure}[H]
    \centering
    \includegraphics[width=\columnwidth]{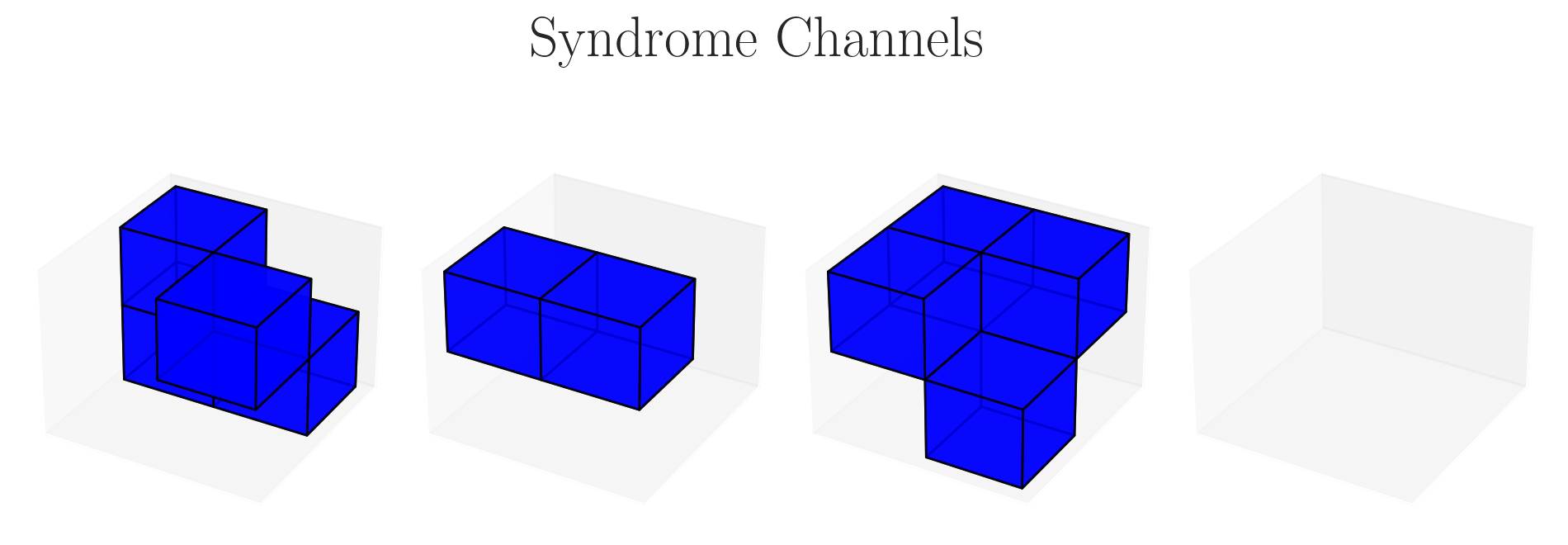}
    \caption{Syndrome Realization}
    \label{fig:syn_vox}
\end{figure}

\begin{figure*}
    \centering
    \includegraphics[width=\textwidth]{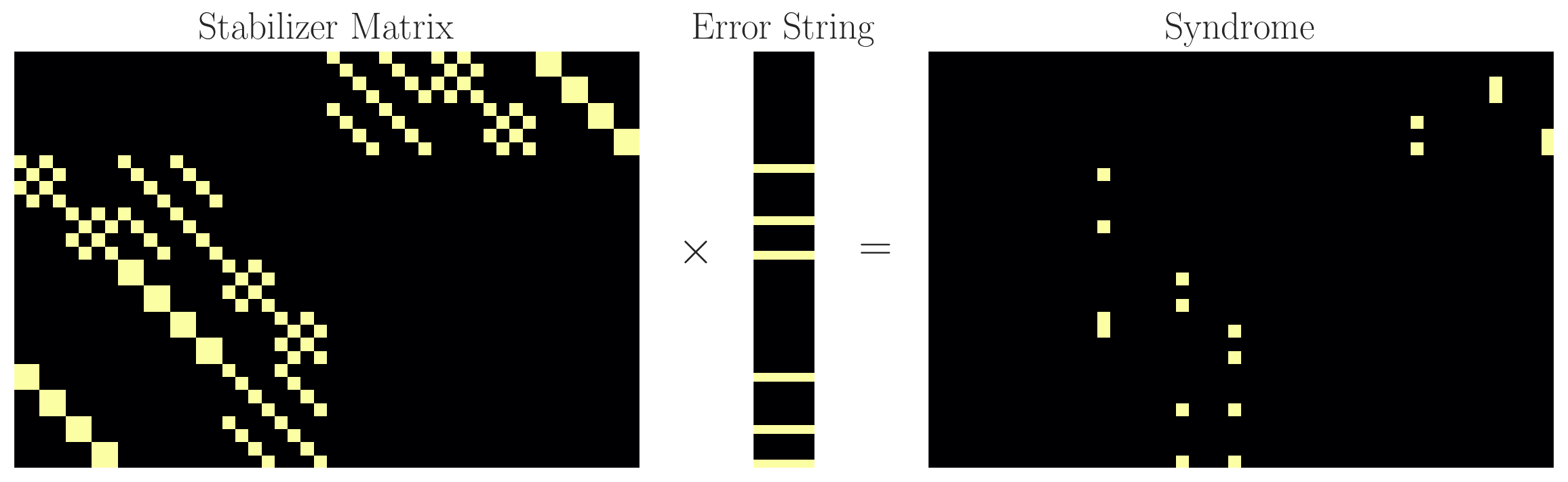}
    \caption{Syndrome Generation}
    \label{fig:syn_gen}
\end{figure*}

We generate an error $E$ with error rate $p = 0.1$ and apply it to the stabilizer matrix $H_{t3d}(L)$ as follows: $\sigma = H_{t3d}(L) \times E$. This operation is shown in Fig. \ref{fig:syn_gen}, and results in a syndrome $\sigma$, which can be split into four channels, as shown in Fig. \ref{fig:syn_vox}.

\subsection{Neural Decoder}
\begin{figure}[H]
    \centering
    \includegraphics[scale=0.175]{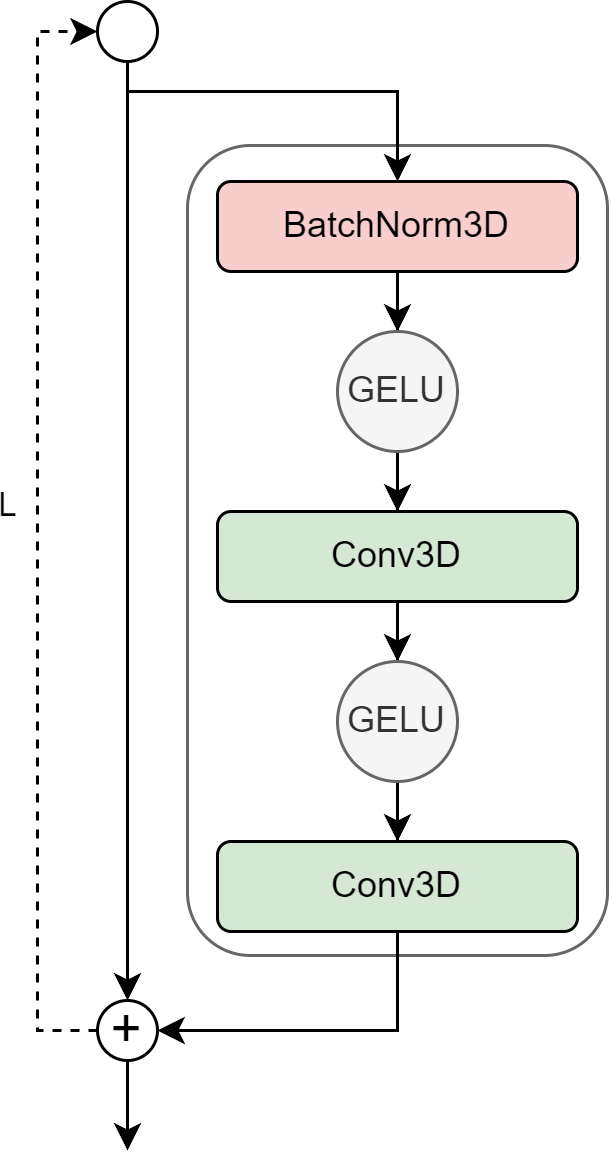}
    \caption{WideResBlock Architecture}
    \label{fig:wideblock}
\end{figure}
\begin{figure}[h]
    \centering
    \includegraphics[height=0.45\textheight]{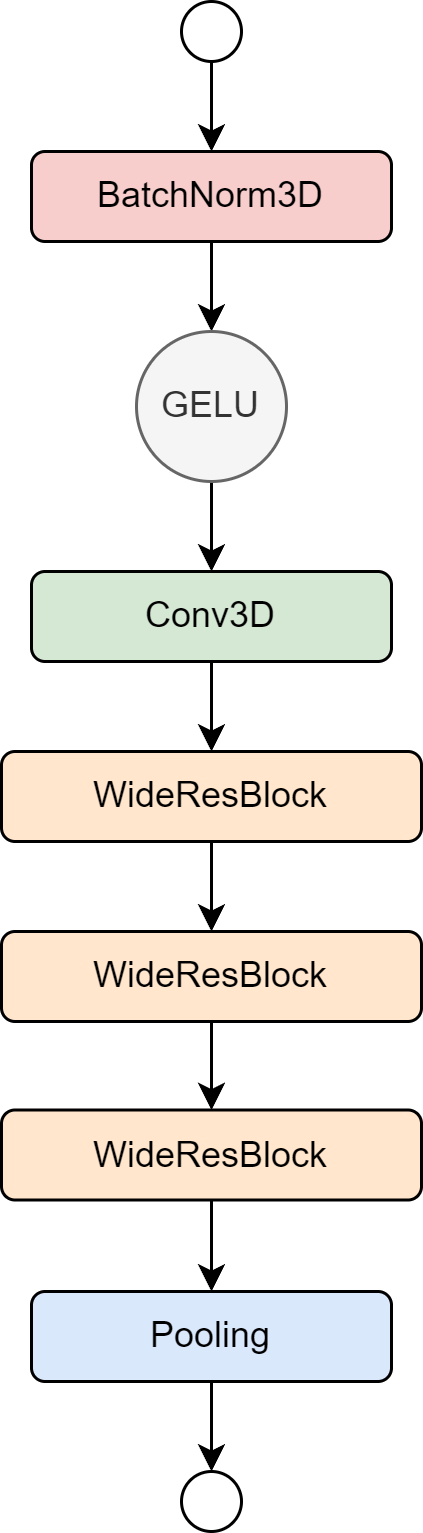}
    \caption{Network Architecture}
    \label{fig:wideresnet}
\end{figure}

\section{Results}\label{sec:app_res}
\begin{figure}[H]
    \centering
    \includegraphics[width=\columnwidth]{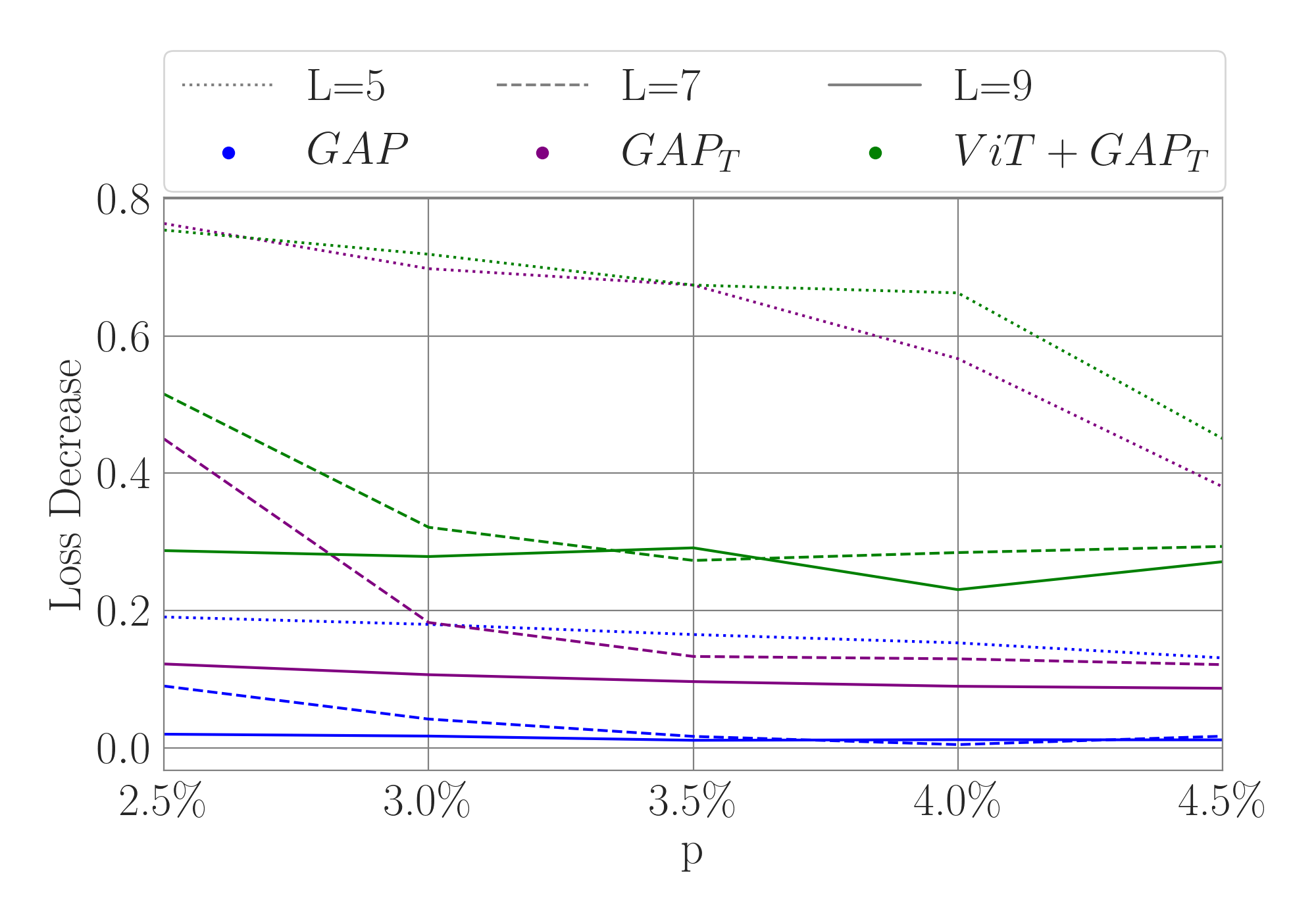}
    \caption{Trainability Plot}
    \label{fig:train2d}
\end{figure}

Here in Fig.\ref{fig:train2d} the trainability surface shown in Fig.\ref{fig:trainability} is illustrated as curves.

\section{Discussion}
\begin{table}[H]
\caption{System Sepcs used for the experiments.}
\begin{tabular}{lll}
\hline
                      & \textit{\textbf{System 1}} & \textit{\textbf{System 2}} \\ \hline
\textit{\textbf{CPU}} & AMD Ryzen 1400 & 2x Intel Xeon Platinum 8360Y \\
\textit{\textbf{GPU}} & NVIDIA RTX 2060            & 4x NVIDIA A100             \\
\textit{\textbf{RAM}} & 2x 8GB 2400MHz             & 16x 32GB 3200MHz           \\ \hline
\end{tabular}\label{tab:specs}
\end{table}
\end{appendices}
\end{document}